\documentclass{raa}
\usepackage{graphicx,times}
\usepackage{natbib}
\usepackage{amssymb,amsmath}
\bibpunct{(}{)}{;}{a}{}{,}

\usepackage[a4paper=true,dvipdfm=true,pagebackref=true]{hyperref}
\hypersetup{pdftitle = The title of my PDF, pdfauthor = My name, pdfsubject= The subject, pdfkeywords = keyword1 keyword2 keyword3} 
\hypersetup{colorlinks = true, linkcolor = green, anchorcolor = red, citecolor = blue, filecolor = red, pagecolor = red, urlcolor = red}

\begin{document}

\title{Optical Spectroscopy and Photometry of Main-Belt Asteroids with a High Orbital
Inclination}

 \volnopage{ {\bf 2015} Vol.\ {\bf X} No. {\bf XX}, 000--000}
 \setcounter{page}{1}

\author{Aya Iwai\inst{1}, Yoichi Itoh\inst{2}, Tsuyoshi Terai\inst{3},
Ranjan Gupta\inst{4}, Asoke Sen\inst{5}, Jun Takahashi\inst{2}}

   \institute{National Institute of Advanced Industrial Science and Technology,
1-1-1, Umezono, Tsukuba, Ibaraki 305-8563, Japan \\
        \and
Nishi-Harima Astronomical Observatory, Center for Astronomy, 
University of Hyogo, 
407-2, Nishigaichi, Sayo, Hyogo 679-5313, Japan; 
             {\it yitoh@nhao.jp} \\
        \and
National Astronomical Observatory of Japan,
650 North Aohoku Place, Hilo Hawaii 96720, USA \\
        \and
Inter-University Centre for Astronomy and Astrophysics,
Ganeshkhind, Pune 411 007, India \\
        \and
Department of Physics, Assam University,
Silchar 788 001, India \\
\vs \no
   {\small Received 2015 xxxx xx; accepted 2015 xxxx xx}
}

\abstract{
We carried out low-resolution optical spectroscopy of 51 main-belt
asteroids, most of which have a highly-inclined orbit.
They are selected from the D-type candidates in the SDSS-MOC 4 catalog.
Using the University of Hawaii 2.2 m telescope and the Inter-University
Centre for Astronomy and Astrophysics 2 m telescope in India,
we determined the spectral types of 38 asteroids.
Among them, eight asteroids were classified as D-type asteroids.
Fractions of D-type asteroids are $3.0\pm1.1$ \% for low orbital inclination
main-belt asteroids and $7.3\pm2.0$ \% for high orbital inclination 
main-belt asteroids.
We consider that a part of
D-type asteroids were formed within the ecliptic region between the main belt and Jupiter, 
and were then perturbated by Jupiter.
}

   \authorrunning{A. Iwai and Y. Itoh}            
   \titlerunning{Highly-Inclined Asteroids}  
   \maketitle

%
\section{Introduction}          
\label{sect:intro}


Asteroids with a high orbital inclination are outlier.
\cite{Terai11} surveyed small asteroids at high ecliptic latitudes.
By using wide and deep optical images taken with Suprime-Cam mounted on the
Subaru Telescope, they detected 656 asteroids with a high orbital
inclination.
The orbital semi-major axes of
most of the asteroids were derived to be between 2.0 astronomical units (AU)
and 3.3 AU.
\cite{Terai13} found that the cumulative size distribution of
the high orbital inclination main-belt asteroids (hereafter MBAs)
was shallower than that of the low orbital inclination MBAs.
It is considered that MBAs with a high orbital inclination
have high collisional velocities.
The shallow size distribution for high orbital inclination MBAs
indicates that a large body has a higher disruptive strength under 
hyper-velocity impacts.

The size distribution of high orbital inclination asteroids 
provides us with information 
regarding the process of planetesimal collision under 
different conditions from those of low orbital inclination planetesimals.
However, the origin of high orbital
inclination asteroids is still uncertain.
\cite{Nagasawa} calculated the orbital evolution of asteroids.
They considered the depletion of gas in the Solar Nebula.
An inner edge of the gas was moving outward at a velocity of
$1\times10^{-5}$ AU yr$^{-1}$.
It was revealed that the eccentricity and inclination of the asteroids
increased due to the motion of secular resonances caused by the gas
depletion, while semi-major axis of the asteroids did not change.
\cite{Ida} investigated 
the orbital evolution of planetesimals near to a proto-planet.
Their numerical simulations showed that the planetesimals moved inward
or outward with increasing orbital inclinations and/or
eccentricities due to the gravitational
perturbation of the proto-planet.

In this paper, we focus on D-type asteroids.
D-type asteroids are abundant in the Trojan and main-belt population 
with the semi-major axis beyond $\simeq$ 3 AU.
Optical reflectance of the D-type asteroids increases significantly with
wavelengths.
Such optical reflectance and low albedo of the D-type asteroids are similar
to those of some comet nuclei.
Many properties of the D-type asteroids, including its spatial distribution
and optical reflectance, have been discussed in the context of cometary
evolution and asteroid populations.
Detail discussion on the D-type asteroids are found, 
e.g., in \cite{Fitzsimmons}.

We conducted low-resolution optical spectroscopy of main-belt asteroids.
Most of these asteroids have a highly-inclined orbit and optical colors
consistent with spectral type D.
Comparison of D-type ratio in high and low inclination populations 
will help us to understand the formation and evolution process of 
high inclination asteroids.
However, we cannot directly apply the D-type ratio from large volume
photometric catalogs, e.g., the SDSS-MOC 4 catalog,
because the classification using multi-color photometry is not certain
(\citealt{Carvano}, \citealt{DeMeo2013}, \citealt{DeMeo2014}).
Therefore we carried out the spectroscopic observations of confirming
the D-type candidates and estimating the real D-type ratio.
Hereafter we define low orbital inclination MBAs as the asteroids with
the semi-major axis between 2.1 AU and 3.3 AU and with the inclination less
than 10 \degr, and high orbital inclination MBAs as the asteroids
with the same semi-major axis range with the inclination equal to or
larger than 10 \degr.

\section{Observations}

We carried out low-resolution optical spectroscopy of MBAs using
the Wide Field Grism Spectrograph 2 (WFGS2) mounted on the 
University of Hawaii (UH) 2.2 m telescope.
The data were obtained on October 30 and 31, 2008 and
October 19 and 20, 2011 with the low-dispersion grism 1 and
the 1\farcs4 width slit. 
These instrument settings achieved a wavelength coverage of 440 -- 830
nm and a spectral resolution of $\sim$ 410 at 650 nm.
Slit orientations have been fixed in the north-south direction.
The telescope was operated in the non-sidereal tracking mode.
The integration time for each object was between 180 s and 600 s.
Facilities of Inter-University Centre for Astronomy and
Astrophysics (IUCAA) were also used.
We employed the IUCAA Faint Object Spectrograph and Camera (IFOSC) 
mounted on the 2 m telescope
at the IUCAA Girawali Observatory (IGO), India.
The data were obtained on December 28 and 29, 2008
with the IFOSC 5 grism and
the 1\farcs5 width slit. 
The wavelength coverage was 520 -- 1030
nm with a spectral resolution of $\sim$ 650 at 650 nm.
Slit orientations have been fixed in the north-south direction.
The telescope was operated in the non-sidereal tracking mode.
The integration time for each object was between 300 s and 600 s.

Several criteria were used to select the targets.
First, we selected asteroids with an orbital semi-major axis of between 2.1 
AU and 3.3 AU, and with an orbital inclination
greater than 10 \degr. 
We selected high orbital inclination asteroids because lack of the high
inclination samples in the previous studies.
One may consider that definition of the high and low orbital inclination
asteroids should be based on the $\nu_{6}$ resonance.
Because the orbital inclination of the $\nu_{6}$ resonance varies
with the semi-major axis,
classification based on the $\nu_{6}$ resonance may complicate the discussion on
the selection bias of the targets.
Instead, we simply define the high and low orbital inclination
asteroids based on their inclinations.
Most of the selected asteroids are listed in the SDSS-MOC 4 catalog and
their optical magnitudes are given.
We constructed an optical color-color diagram of the selected asteroids
using the SDSS magnitudes,
and classified those objects with $g - r \geq 0.5$ mag and $r - i \geq 0.2$ mag
as candidates for D-type asteroids (\citealt{Ivezic}).
Because we selected the targets on the $gri$ color-color diagram,
we considered the selection bias also on the $gri$ color-color diagram
(section 4).

We were not able to completely fill the observing time
whilst searching for D-type asteroids.
Therefore, when we could not find any suitable candidates, we
observed high orbital inclination asteroids instead,
even if their SDSS colors were unknown.
As a result, we observed a total of 48 high orbital
inclination
asteroids.
We also observed three low orbital inclination asteroids for reference.
The optical spectra of dwarfs with a spectral type of G were also taken as
spectral standards.

After the observations of this work, 
\cite{DeMeo2013} proposed a new method for the taxonomy with using
the SDSS $g$-,$r$-,$i$-, and $z$-band filters.
They claimed that the asteroid types are identified with the $gri$
slope and $i-z$ color.
The D-type asteroids are classified as the group with the reddest $gri$ slope
and the reddest $i-z$ color.
The targets in this paper were classified into the D-type, L-type, 
or the S-type with the large
$i-z$ magnitudes by the classification of \cite{DeMeo2013}.
Efficiency of the discovery of the D-type asteroids would increase,
if we used the classification method of \cite{DeMeo2013}.
This will be attempted in future.

Additional photometric observations were carried out on
November 7, 8, and 9, 2012 with the Multiband Imager for Nayuta Telescope (MINT)
mounted on the Nayuta Telescope at Nishi-Harima Observatory, Japan.
We observed 4 asteroids whose SDSS magnitudes were unknown.
We used SDSS $g-$, $r-$, and $i-$band filters.
The field of view was 10 \arcmin $\times$ 10 \arcmin with a pixel
scale of 0\farcs3.
The full width at half maximum of the point spread functions
was typically 2\farcs0
or 2\farcs5.
The integration time was between 30 s and 600 s for each object.
Since an asteroid rotates and its brightness may vary over time,
we repeatedly observed an asteroid in the $r-$band.
The objects we investigated are summarized in table \ref{target}.

\begin{table}
\begin{center}
\caption{Targets and Assigned Spectral Type}
\label{target}
\begin{tabular}{lllllllccc}
\hline
\hline
Asteroid number & $a$ [AU] & $i$ [$\degr$] & 
APmag$^{\amalg}$ & UT Date & sec $z$ & Ref. star (sec $z$) & Telescope & 
\multicolumn{2}{c}{Spectral Type} \\
 & & & & & & & &this work & previous work \\
\hline
171 & 3.13 & 2.55 & 13.61 & 2008/10/31 & 1.07 & SAO 128811 (1.08) & UH & C & C$^{\dagger}$\\
556 & 2.46 & 5.24 & 12.56 & 2008/10/31 & 1.03 & SAO 75211 (1.00) & UH & S & S$^{\dagger}$\\
729 & 2.76 & 18.00 & 14.27 & 2011/10/20 & 1.07 & SAO 111802 (1.03) & UH  & D & D$^{\dagger}$\\
1241 & 3.19 & 23.55 & 14.53 & 2008/11/01 & 1.16 & SAO 109922 (1.04) & UH & C \\
1264 & 2.86 & 25.00 & 14.09 & 2008/12/28 & 1.66 & SAO 41806 (1.74) & IUCAA & C & C$^{\dagger}$\\
1266 & 3.37 & 17.10 & 14.59 & 2011/10/20 & 1.03 & SAO 75098 (1.03) & UH & D & D$^{\ddagger}$\\
1406 & 2.70 & 12.40 & 14.88 & 2011/10/21 & 1.09 & SAO 145806 (1.19) & UH & & Ld$^{\dagger}$\\
2023 & 2.88 & 22.40 & 14.80 & 2011/10/21 & 1.15 & SAO 38934 (1.12) & UH & X \\
3139 & 3.20 & 20.53 & 15.52 & 2008/11/01 & 1.18 & SAO 73774 (1.30) & UH & C & C$^{\ddagger}$\\
3346 & 3.19 & 21.50 & 15.60 & 2008/11/01 & 1.11 & SAO 109922 (1.04) & UH & C \\
4423 & 3.39 & 19.30 & 16.80 & 2008/11/01 & 1.26 & SAO 111908 (1.25) & UH & D & B$^{\dagger}$\\
5222 & 2.77 & 34.58 & 15.69 & 2008/11/01 & 1.39 & SAO 111908 (1.25) & UH & C \\
5889 & 3.05 & 19.18 & 16.88 & 2008/11/01 & 1.12 & SAO 130080 (1.07) & UH & X \\
6052 & 3.24 & 21.70 & 15.88 & 2008/12/29 & 1.80 & SAO 115381 (1.32) & IUCAA & C \\
6148 & 2.28 & 22.70 & 17.77 & 2011/10/21 & 1.12 & SAO 145806 (1.19) & UH \\
7290 & 2.56 & 24.76 & 15.13 & 2008/11/01 & 1.02 & SAO 109922 (1.04) & UH & C \\
7331 & 3.12 & 22.35 & 15.88 & 2008/10/31 & 1.08 & SAO 75211  (1.00) & UH \\
8250 & 3.13 & 17.05 & 16.60 & 2008/11/01 & 1.10 & SAO 109922 (1.04) & UH & S \\
8882 & 2.36 & 23.10 & 16.24 & 2011/10/21 & 1.13 & SAO 145806 (1.19) & UH & D \\
13617 & 2.65 & 22.10 & 17.56 & 2011/10/21 & 1.23 & SAO 38934 (1.12) & UH & S \\
13810 & 3.15 & 17.04 & 16.81 & 2008/11/01 & 1.06 & SAO 130080 (1.07) & UH & D \\
15213 & 3.14 & 21.54 & 17.50 & 2008/10/31 & 1.11 & SAO 93158  (1.08) & UH & C \\
15410 & 3.11 & 17.15 & 16.91 & 2008/11/01 & 1.26 & SAO 111908 (1.25) & UH & C \\
15976 & 2.98 & 10.00 & 16.94 & 2011/10/20 & 1.03 & SAO 75098  (1.03) & UH & S \\
17435 & 2.37 & 22.10 & 17.58 & 2011/10/21 & 1.44 & SAO 130166 (1.52) & UH \\
21552 & 2.60 & 22.10 & 17.34 & 2011/10/21 & 1.24 & SAO 130166 (1.52) & UH \\
25572 & 2.59 & 22.90 & 17.00 & 2011/10/20 & 1.08 & SAO 111802 (1.03) & UH & S \\
25850 & 3.12 & 17.65 & 16.90 & 2008/11/01 & 1.10 & SAO 130080 (1.07) & UH & D \\
26571 & 2.79 & 15.20 & 17.47 & 2011/10/20 & 1.42 & SAO 75098  (1.03) & UH \\
33750 & 2.80 & 32.70 & 17.15 & 2008/10/31 & 1.15 & SAO 128811 (1.08) & UH & C \\
37977 & 3.13 & 20.17 & 15.84 & 2011/10/21 & 1.11 & SAO 145806 (1.19) & UH & D \\
43142 & 2.59 & 14.40 & 16.13 & 2011/10/21 & 1.22 & SAO 147902 (1.27) & UH & S \\
43815 & 3.17 & 24.11 & 15.23 & 2008/11/01 & 1.08 & SAO 109922 (1.04) & UH & X \\
44215 & 2.58 & 12.40 & 17.22 & 2011/10/20 & 1.03 & SAO 75098  (1.03) & UH \\
45411 & 2.63 & 12.10 & 16.56 & 2011/10/21 & 1.16 & SAO 38934  (1.12) & UH & S \\
46564 & 3.19 & 17.17 & 17.05 & 2008/10/31 & 1.12 & SAO 128811 (1.08) & UH & X \\
47334 & 2.63 & 13.80 & 16.28 & 2011/10/20 & 1.04 & SAO 75098  (1.03) & UH \\
49591 & 3.22 & 17.68 & 16.40 & 2008/11/01 & 1.05 & SAO 109922 (1.04) & UH & C \\
54750 & 3.22 & 17.26 & 16.85 & 2008/11/01 & 1.25 & SAO 111908 (1.25) & UH & C \\
57036 & 2.34 & 23.80 & 17.76 & 2011/10/20 & 1.22 & SAO 111802 (1.27) & UH & S \\
64969 & 2.29 & 24.30 & 16.28 & 2011/10/21 & 1.30 & SAO 145806 (1.19) & UH & D \\
70104 & 3.17 & 22.95 & 16.70 & 2008/11/01 & 1.31 & SAO 111908 (1.25) & UH & S \\
73506 & 3.20 & 18.33 & 16.69 & 2008/10/31 & 1.10 & SAO 93158  (1.08) & UH \\
\hline
\end{tabular}
\end{center}
\end{table}

\begin{table}
\begin{center}
\begin{tabular}{lllllllccc}
\hline
\hline
Asteroid number & $a$ [AU] & $i$ [$\degr$] & 
APmag$^{\amalg}$ & UT Date & sec$z$ & Ref. star (sec $z$) & Telescope & 
\multicolumn{2}{c}{Spectral Type} \\
 & & & & & & & & this work & previous work \\
\hline
93751 & 2.37 & 25.60 & 16.48 & 2011/10/21 & 1.37 & SAO 148719  (1.22) & UH \\
94030 & 2.61 & 28.05 & 16.60 & 2008/10/31 & 1.06 & SAO 75211   (1.00) & UH & C \\
99202 & 3.20 & 22.24 & 17.02 & 2008/11/01 & 1.07 & SAO 130080  (1.07) & UH & C \\
163943 & 2.53 & 17.30 & 18.05 & 2011/10/21 & 1.21 & SAO 148719 (1.22) & UH \\
193488 & 2.58 & 20.36 & 16.48 & 2008/11/01 & 1.11 & SAO 75874  (1.07) & UH \\
198575 & 2.66 & 29.47 & 17.16 & 2008/10/31 & 1.03 & SAO 75211  (1.00) & UH & C \\
213177 & 2.38 & 24.20 & 17.80 & 2011/10/20 & 1.02 & SAO 75438  (1.00) & UH \\
307783 & 3.09 & 17.73 & 16.54 & 2008/11/01 & 1.07 & SAO 109922 (1.04) & UH & X \\
\hline
\end{tabular}
\end{center}
$^{\amalg}$: APmag is calculated on the JPL small body Database browser. \\
$^{\dagger}$: \cite{Busb} \\
$^{\ddagger}$: \cite{Lazzaro}
\end{table}

We reduced the spectroscopic data with the following steps: 
overscan subtraction,
bias subtraction, flat fielding, removal of scattered light
caused by the telescope and/or the instrument, extraction
of a spectrum, and wavelength calibration.
For the wavelength calibration, we used the O I and Na I lines of the sky for
the UH data, and the emission lines of the Ne-Ar lamp for the IUCAA data.
The spectra of the asteroids were then divided by the dwarf spectra.
Thus, the spectra indicate the reflectivity of the asteroid.
The spectrum of each asteroid was normalized to unity at 550 nm.

We reduced the photometric data with the following steps: 
overscan subtraction,
bias subtraction, and flat fielding with twilight flat frames.
We used aperture photometry to measure the fluxes of the asteroids
and several background stars imaged in the same.
The aperture radius varied between 6 and 12 pixels,
depending on the seeing size.
Magnitudes of the asteroid were calculated using relative photometry
with those of the background stars (SDSS DR9 catalog; \citealt{Ahn}).
We used IRAF packages for all data-processing.

\section{Results}

Figure \ref{spec} shows the spectra of the asteroids and
table \ref{photometry} lists their magnitudes and colors.
The signal-to-noise ratios of the spectra in wavelengths
between 660 nm and 740 nm ranged from 20 to 350 with
a median value of 60.

\cite{Busa} and \cite{Busb} defined 26 asteroid spectral types.
From the low-resolution optical spectra of thousands of asteroids,
they calculated the average reflectivities of nine wavelength regions
between 435 nm and 925 nm for each spectral type.
We made temperate spectra of C-, S-, X-, D-, and V-type asteroids
by interpolating the average reflectivities
for each asteroid spectral type.
The temperate spectra were then normalized to unity at 550 nm.
The template spectrum of a C-type asteroid is almost constant,
whereas the spectra of S-type and V-type asteroids
peak at approximately 700 nm.
The template spectrum of an X-type asteroid increases gradually with wavelength,
and that of a D-type asteroid increases significantly with wavelength.
Individual asteroid spectra differ even if the asteroids are
classified into the same spectral type.
The standard deviation of the spectra among the
same spectral type of asteroids is approximately 0.02.
We calculated the residuals between the observed spectra and
five template spectra.
We then assigned a spectral type to an asteroid if the residuals were 
at a minimum
and were smaller than 0.07.
As a result, we classified 38 asteroids; 9 asteroids were classified
as S-type, 16 as C-type, 5 as X-type, 8 as D-type, and none as V-type.
Among them, the spectral types of six asteroids have been previously assigned
in \cite{Busb}
and the spectral type of one asteroid was assigned in \cite{Lazzaro}.
The spectral-type assignments in this work are consistent with the previous 
assignments for 6 assignments out of 7 assignments.
\cite{Carvano} assigned spectral types for 63468 asteroids based on the
SDSS photometric data.
Among them, 22019 asteroids have two or more observations.
Of these, 14962 show taxonomic variations.
They employed 9 classes for the spectral assignments,
and found the transition between dissimilar classes, e.g., D to S, X to S,
X to S, and C to S.
They proposed extreme space weathering, contamination by metal, the coexistence
of different mineralogies in the same body, or phase reddening.
We consider that one inconsistent assignment between this work and the 
previous study may be attributed to inhomogeneity of the asteroid
surface, or ambiguity on the assignments.


\begin{figure}
 \centering
 \includegraphics[width=14.0cm, angle=0]{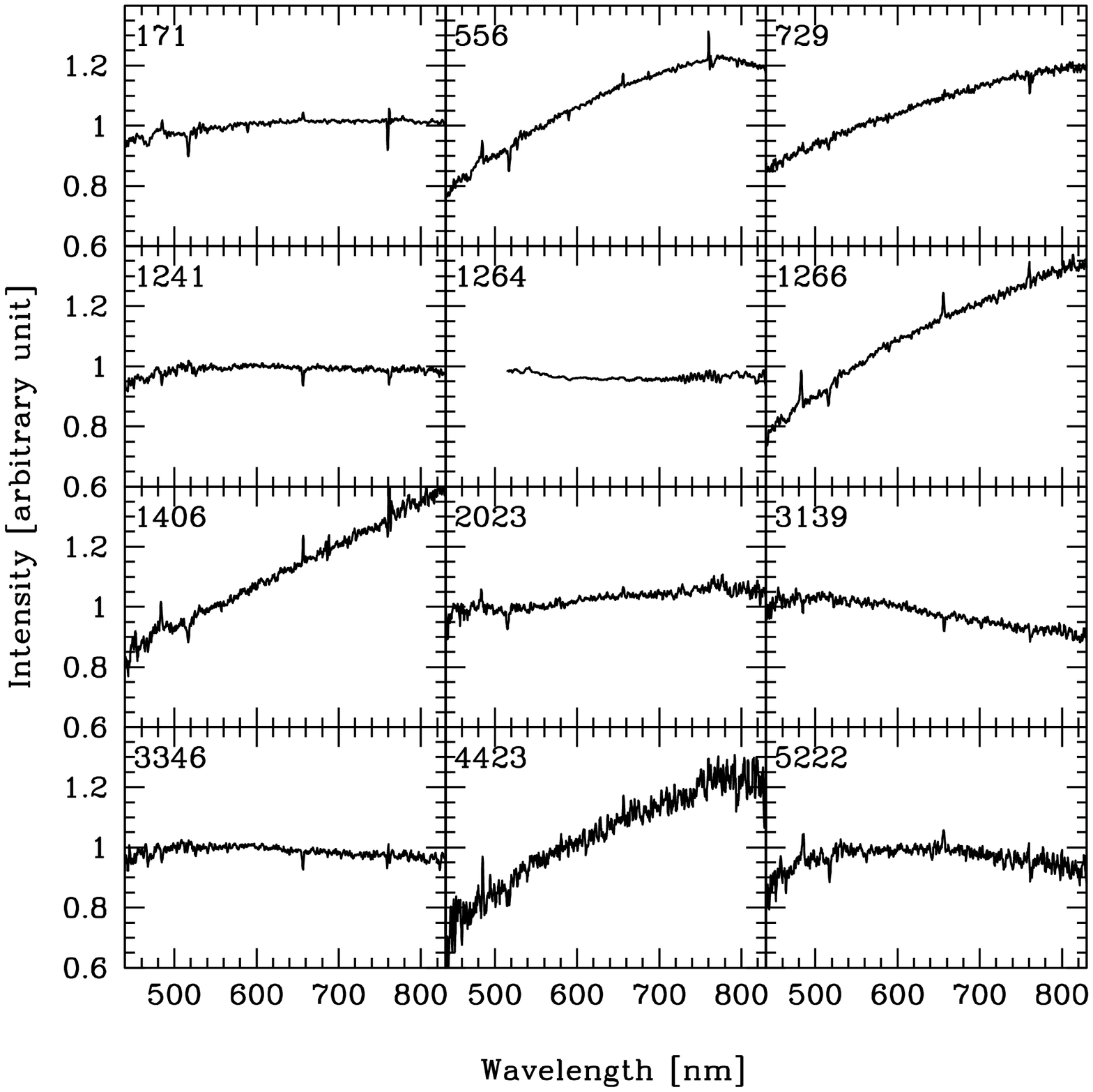}
 \caption{Optical spectra of the high orbital inclination asteroids.
The reflectance are normalized to unity at 550 nm.
Asteroids 171, 556, and 15123 are low orbital inclination asteroids.
Spike patterns appear in the spectra are artifacts due to
telluric absorptions and photospheric absorptions of the standard
stars.}
 \label{spec}
\addtocounter{figure}{-1}
\end{figure}

\begin{figure}
 \centering
 \includegraphics[width=14.0cm, angle=0]{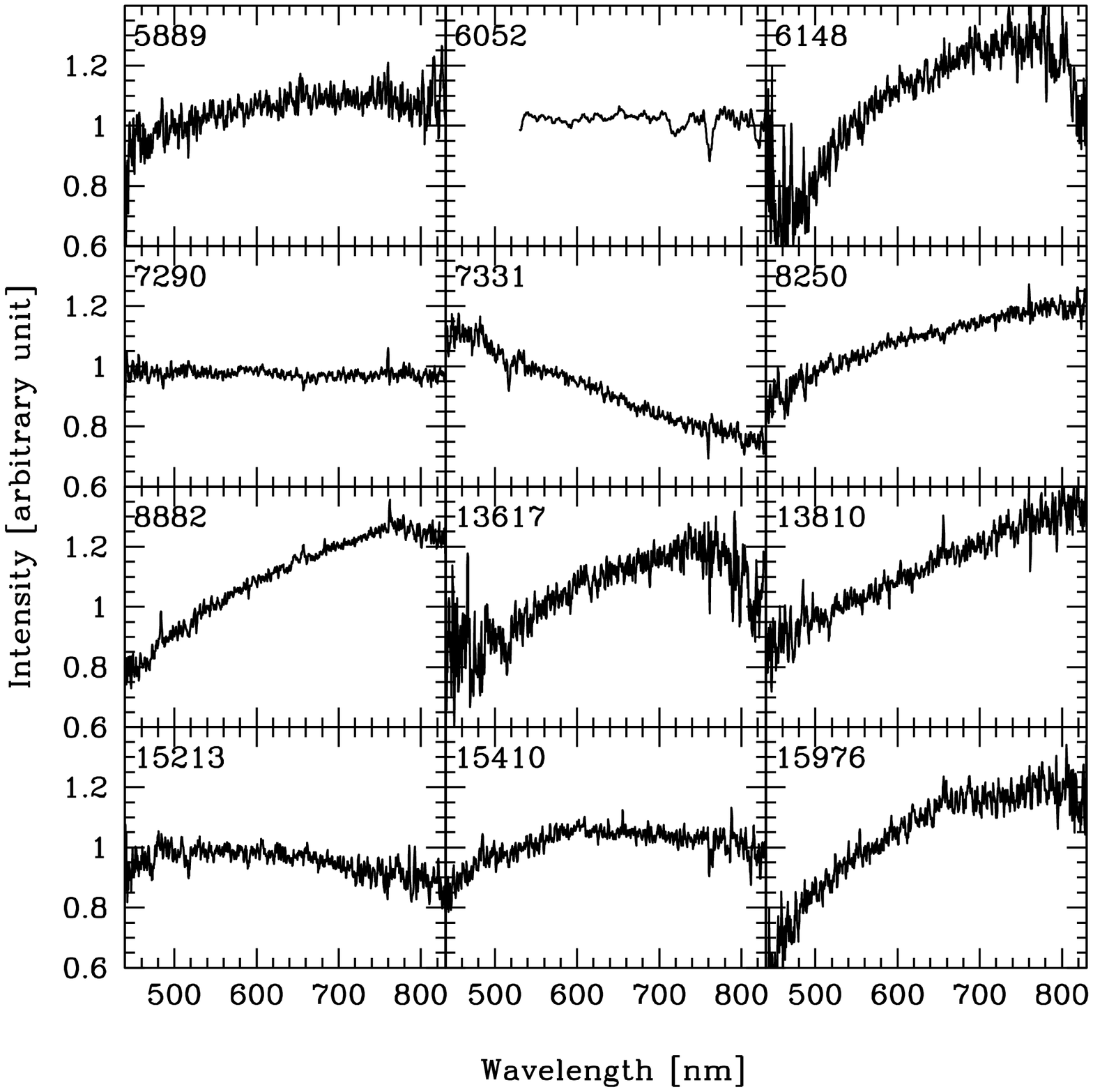}
 \caption{continued}
 \addtocounter{figure}{-1}
\end{figure}

\begin{figure}
 \centering
 \includegraphics[width=14.0cm, angle=0]{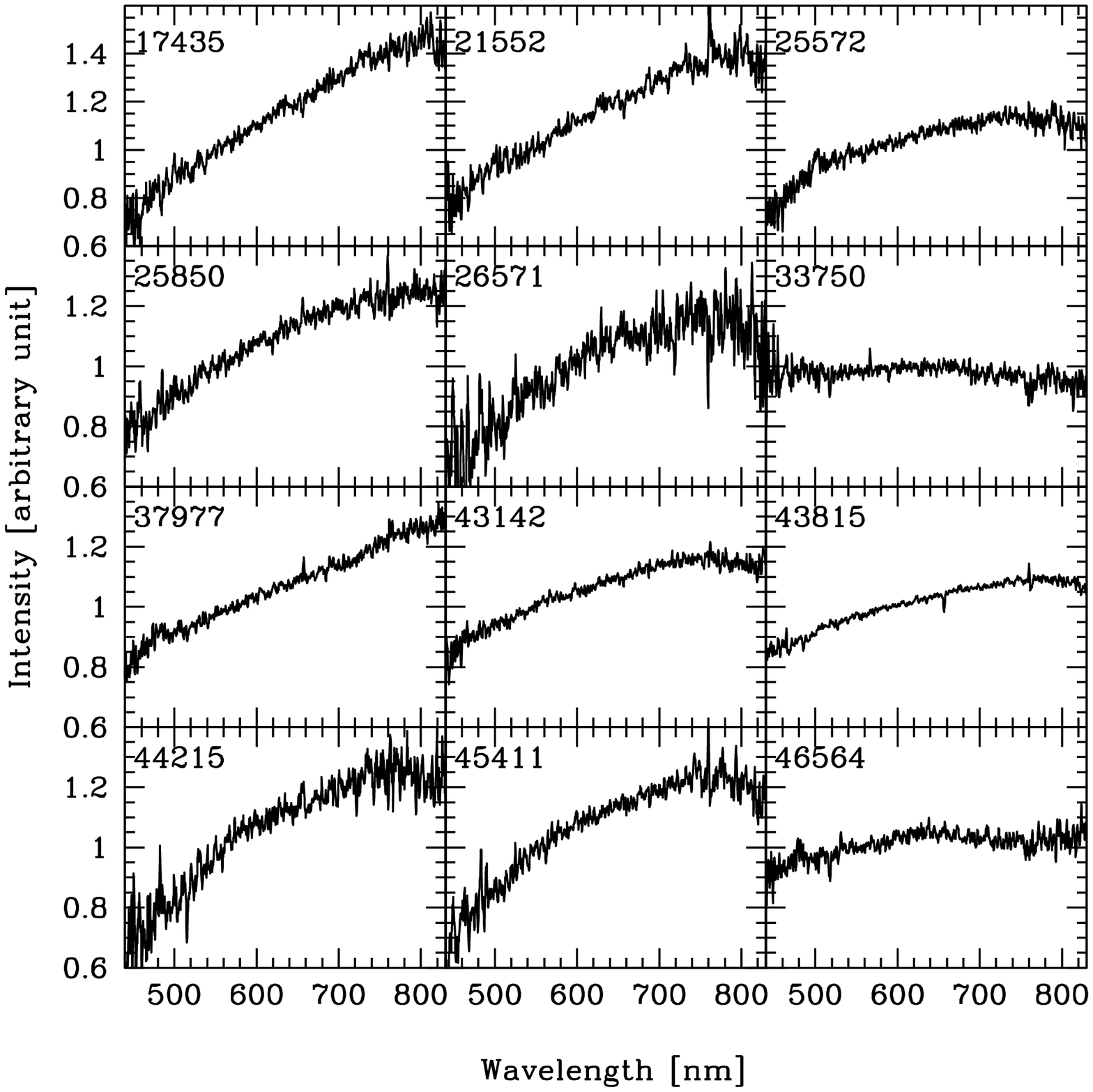}
 \caption{continued}
 \addtocounter{figure}{-1}
\end{figure}

\begin{figure}
 \centering
 \includegraphics[width=14.0cm, angle=0]{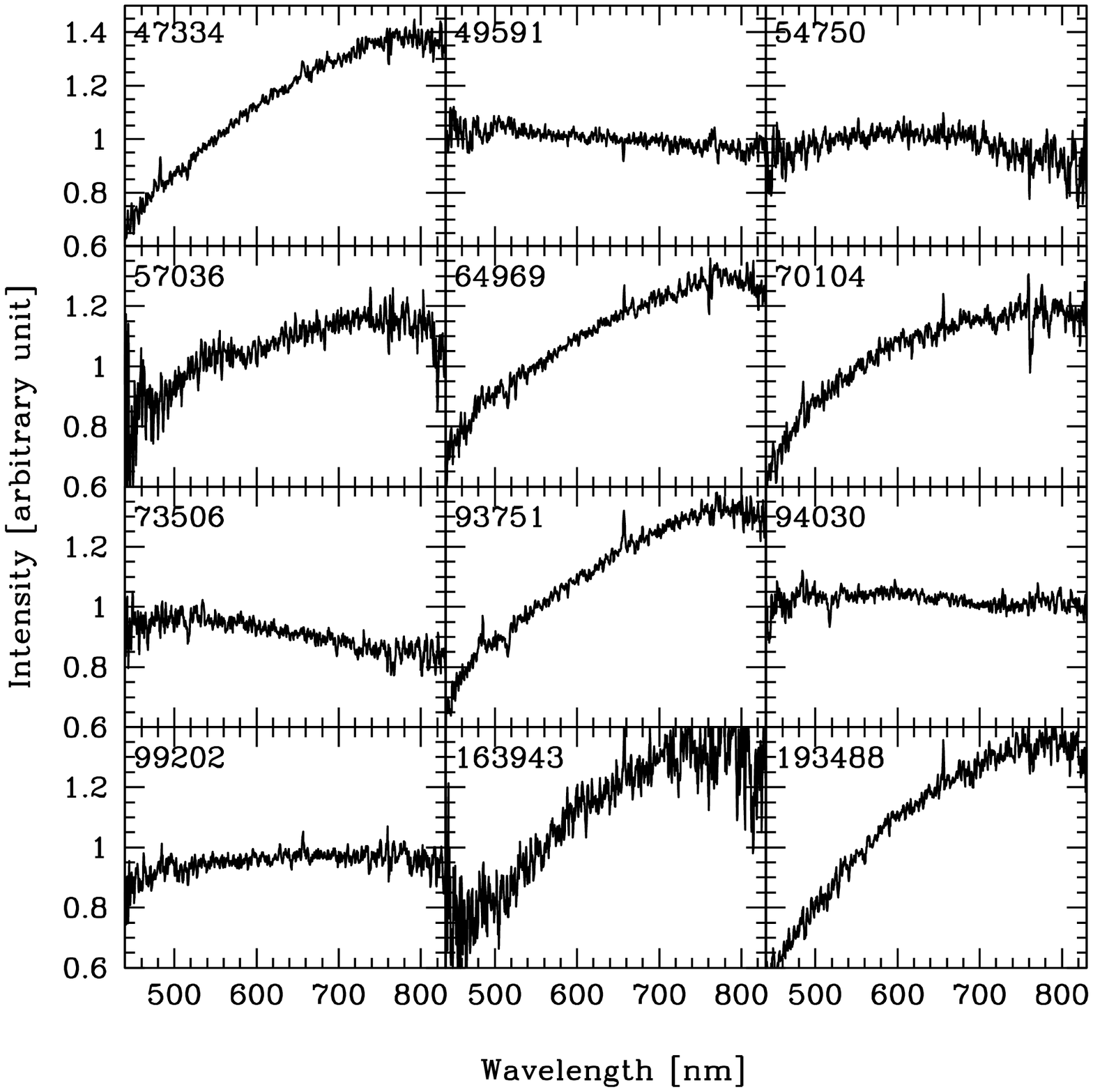}
 \caption{continued}
 \addtocounter{figure}{-1}
\end{figure}

\begin{figure}
 \centering
 \includegraphics[width=14.0cm, angle=0]{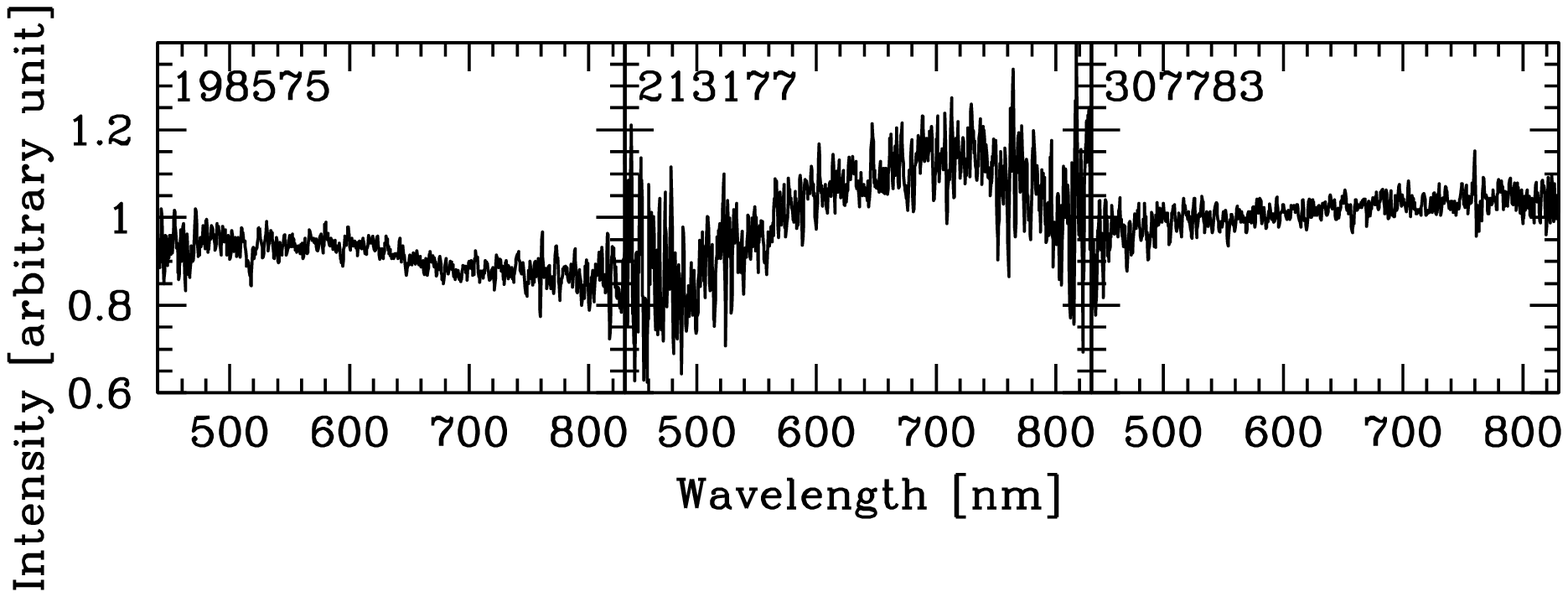}
 \caption{continued}
\end{figure}

\begin{table}
\begin{center}
\caption{Magnitudes and colors of the Asteroids}
\label{photometry}
\begin{tabular}{llll}
\hline
\hline
Asteroid number & $g$-mag & $g-r$ [mag] & $r-i$ [mag] \\
\hline
729   & 15.55$\pm$0.07 & 0.89$\pm$0.09 & 0.19$\pm$0.09 \\
5222  & 16.54$\pm$0.04 & 0.75$\pm$0.10 & $-0.04\pm0.06$ \\
7290  & 17.40$\pm$0.05 & 0.54$\pm$0.07 & 0.17$\pm$0.07 \\
94030 & 17.88$\pm$0.09 & 0.53$\pm$0.13 & 0.15$\pm$0.08 \\
\hline
\end{tabular}
\end{center}
\end{table}

\section{Discussion}

We discuss the spatial distribution of D-type asteroids.
D-type asteroids are abundant in the ecliptic plane between the main belt
and the orbit of Jupiter ($\sim$ 50\%, \citealt{DeMeo2013}).
The number of asteroids that have been categorized as spectral type D
is too small
for a statistical discussion of the spatial distribution of these asteroids.
We used photometric data from the SDSS-MOC4 catalog,
as well as the results of the spectroscopic studies (\citealt{Busb},
\citealt{Lazzaro}, this work).
Because the selection of our targets was based on the color-color
diagram, we carefully combined the photometric studies and the spectroscopic
studies against the selection bias.
In the SDSS-MOC4 catalog, 67,921 low orbital inclination MBAs are listed
and 31,400 high orbital inclination MBAs are listed (Table \ref{smass}).
\cite{Ivezic} claimed that
the SDSS $g-r$ and $r-i$ colors could be used to classify the spectral types
of asteroids.
We defined candidates for D-type asteroids as the objects with $g-r \geq 0.5$
mag and $r-i \geq 0.2$ mag.
A total of 26,261 low orbital inclination MBAs and 11,786 high
orbital inclination MBAs
have been identified as D-type candidates.
Among them, optical spectra of 
103 low orbital inclination candidates and 67 high orbital inclination
candidates have been obtained (\citealt{Busb},
\citealt{Lazzaro}, this work).
Eight and thirteen objects were spectroscopically assigned as D-type
for low orbital inclination candidates
and high orbital inclination candidates, respectively.
We calculated the fraction of D-type asteroids.
For low orbital inclination MBAs, the fraction of D-type asteroids is
$(26,261/67,921)\cdot(8/103) = 3.0\pm1.1\%$.
For high orbital inclination MBAs, the fraction is
$(11,786/31,400)\cdot(13/67) = 7.3\pm2.0\%$.
In this calculation, the uncertainty of the D-type fraction is
controversial.
The uncertainty involves ambiguity of taxonomy, accuracy of the
orbital inclination,
and many other factors.
We do not fully examine these factors.
Instead, we simply took square root of the number of the asteroids as the
uncertainty of the number of the asteroids.
We conclude that the fraction of D-type asteroids is higher in the
high orbital inclination MBAs than in the low
orbital inclination MBAs.
D-type asteroids are abundant in the ecliptic plane between the main belt
and the orbit of Jupiter.
We propose that a number of 
D-type asteroids were formed near to the ecliptic plane between
the main belt and the orbit of Jupiter, and that a part of D-type
asteroids then migrated inward via 
gravitational scattering by Jupiter and gained a high inclination orbit.
One may consider whether the D-type fraction changes
if we apply the D-type candidates criteria of \cite{DeMeo2013} to our targets.
However, we selected the targets based on the $gri$ color-color diagram
following \cite{Ivezic}.
Among our targets, only 4 asteroids match the D-type criteria of
\cite{DeMeo2013}.
This small number of the sample prevent us from statistical discussion
on the spatial distribution of the D-type asteroids.
\cite{DeMeo2013} deduced that the D-type mass ratio in the main-belt is
$\sim$ 2\%, or $\sim$ 4.4\% if the largest four asteroids are removed.
The D-type number ratio in the main-belt deduced in this work
is $\sim$ 4\%, if combining the high inclination and low inclination samples.
This ratio seems consistent with DeMeo and Carry result of 4.4 \% of mass
in the main-belt belong to the D-type asteroids.
However, \cite{DeMeo2013} assumed that the D-type asteroids have bulk density 
of 1 g cm$^{-3}$, roughly half of the density of the S- and C-type asteroids.
Therefore, one may think that the D-type asteroids number ratio should
be $\sim$ 8\%, which is higher than the result that this work presents.
\cite{DeMeo2013} calculate an average albedo of each asteroid type.
We noticed that the albedo of the D-type asteroids is low among all types
and that they have very low albedo especially in the short wavelengths
in the optical.
This low reflectivity may cause the inconsistency between the number ratio
and the mass ratio.

We consider the false-negative rate of the broad-band photometry D-type
selection. For the low orbital inclination asteroids in the SDSS MOC 4 
catalog, 41660 asteroids were not selected as the D-type candidates, based 
on the color-color diagram. Among this sample, 3281 asteroids are actually 
classified as the D-type asteroids, according to the classification of 
\cite{DeMeo2013}. Thus, the false-negative rate is (3281/41660) = 7.9\%, 
and the "true" D-type fraction of the low orbital inclination asteroids is
(26261/67921)(8/103)+(41660/67921)*7.9\% = 7.8\%.
In the same manner, the false-negative rate of the high orbital inclination 
asteroids is 1752/(31400-11786) = 8.9\%, and the "true" D-type fraction is
(11786/31400)(17/63)+(19614/31400)*8.9\% = 12.8\%.
The false-negative rates are high, so that the rates may effect
the results. However, because the false-negative rates for the low orbital 
inclination asteroids and the high orbital inclination asteroids are similar,
we think that the result of higher D-type ratio in the high orbital inclination
asteroids is still hold.

\begin{table}
\begin{center}
\caption{Fractions of the asteroids in the SMASSII catalog (\citealt{Busb})}
\label{smass}
\begin{tabular}{c|c|cc}
\hline
\hline
Inclination & Type & $2.1 < a \leq 2.5$ AU & $2.5 < a \leq 3.3$ AU \\
\hline
& C & $13.9\pm1.8$ \%  & $38.2\pm2.4$ \% \\
$i < 10$ \degr   &  S & $62.3\pm3.8$ \% & $31.1\pm2.2$ \% \\
& D & \multicolumn{2}{c}{$3.0\pm1.1$ \%}\\
\hline
& C & $15.3\pm3.4$ \% & $34.0\pm2.3$ \% \\
$i \geq 10$ \degr & S & $62.6\pm6.9$ \% & $24.0\pm2.0$ \% \\ 
& D & \multicolumn{2}{c}{$7.3\pm2.0$ \%}\\
\hline
\end{tabular}
\end{center}
\end{table}

We next investigated the spatial distribution of C-type asteroids.
C-type asteroids with a low orbital inclination are mainly distributed in
the outer part of the main belt ($a > 2.5$ AU).
If C-type asteroids were also perturbed by Jupiter,
a fraction of the C-type asteroids should also 
have moved inward and acquired a high
inclination.
We calculated the fraction of C-type asteroids with a semi-major
axis between 2.1 AU and 2.5 AU.
The spectral classifications of \cite{Busb} and
\cite{Lazzaro} were used.
Spectral types were assigned to
438 asteroids with a low orbital inclination
and 131 asteroids with a high orbital inclination.
Among them, 46 low orbital inclination asteroids and 
20 high orbital inclination asteroids
were classified as C-type asteroids.
For simplicity, B-type asteroids were also classified as C-type.
For low orbital inclination asteroids with a semi-major axis of
between 2.1 AU and 2.5
AU, the fraction of C-type asteroids is
$(61/438) = 13.9\pm1.8\%$.
For high orbital inclination asteroids with a semi-major axis of
between 2.1 AU and 2.5
AU, the fraction is
$(20/131) = 15.3\pm3.4\%$.
The difference in the fractions of C-type asteroids between the low
orbital inclination 
asteroids and
the high orbital inclination asteroids is not identified.

We also calculated the fractions of the S-type asteroids.
For the low orbital inclination asteroids with the semi-major axis between
2.1 AU and 2.5 AU, the fraction of the S-type asteroids is 
$(273/438) = 62.3\pm3.8\%$.
For the high orbital inclination asteroids with the semi-major axis between
2.1 AU and 2.5 AU, the fraction of the S-type asteroids is
$(82/131) = 62.6\pm6.9\%$.
Difference between the low orbital inclination asteroids and the high 
orbital inclination asteroids is not identified also for
the S-type asteroids.
We consider that 
the orbits of asteroids located far from Jupiter have not been
significantly perturbed by Jupiter.

\cite{Ida} calculated the orbital evolution of planetesimals perturbed
by a proto-planet.
They indicated that planetesimals near the planet move with conserving
Jacobi energy.
The planetesimal's eccentricity and/or inclination
increases once its semi-major axis increases or decreases.
As a result, the planetesimals are distributed
in a V-shape on the semi-major axis
versus the $(e^{2}+i^{2})^{1/2}$ (called the relative velocity)
plane, with the apex of the V-shape located
at the position of the proto-planet.
The following studies extend numerical calculations of gravitational perturbation
to the case of planetesimals with migrating planets (e.g. \citealt{Gomes}),
or to the case of trans-Neptunian objects (e.g. \citealt{Lykawka}).

It is considered that 
a protoplanetary disk still contains a large amount of gas
during the early stages of planetary formation.
\cite{Adachi} calculated the accretion time of planetesimals.
They found that the accretion time of a 10 km-sized planetesimal onto
the Sun exceeds $10^{9}$ yr.
The radii of the asteroids observed in this work are estimated to be several 
tens of kilometers, with an average radius of 30 km.
Thus, we consider that the gas drag in the protoplanetary disk did not
significantly change the orbital semi-major axis of the asteroids.

\begin{figure}
 \centering
 \includegraphics[width=14.0cm, angle=0]{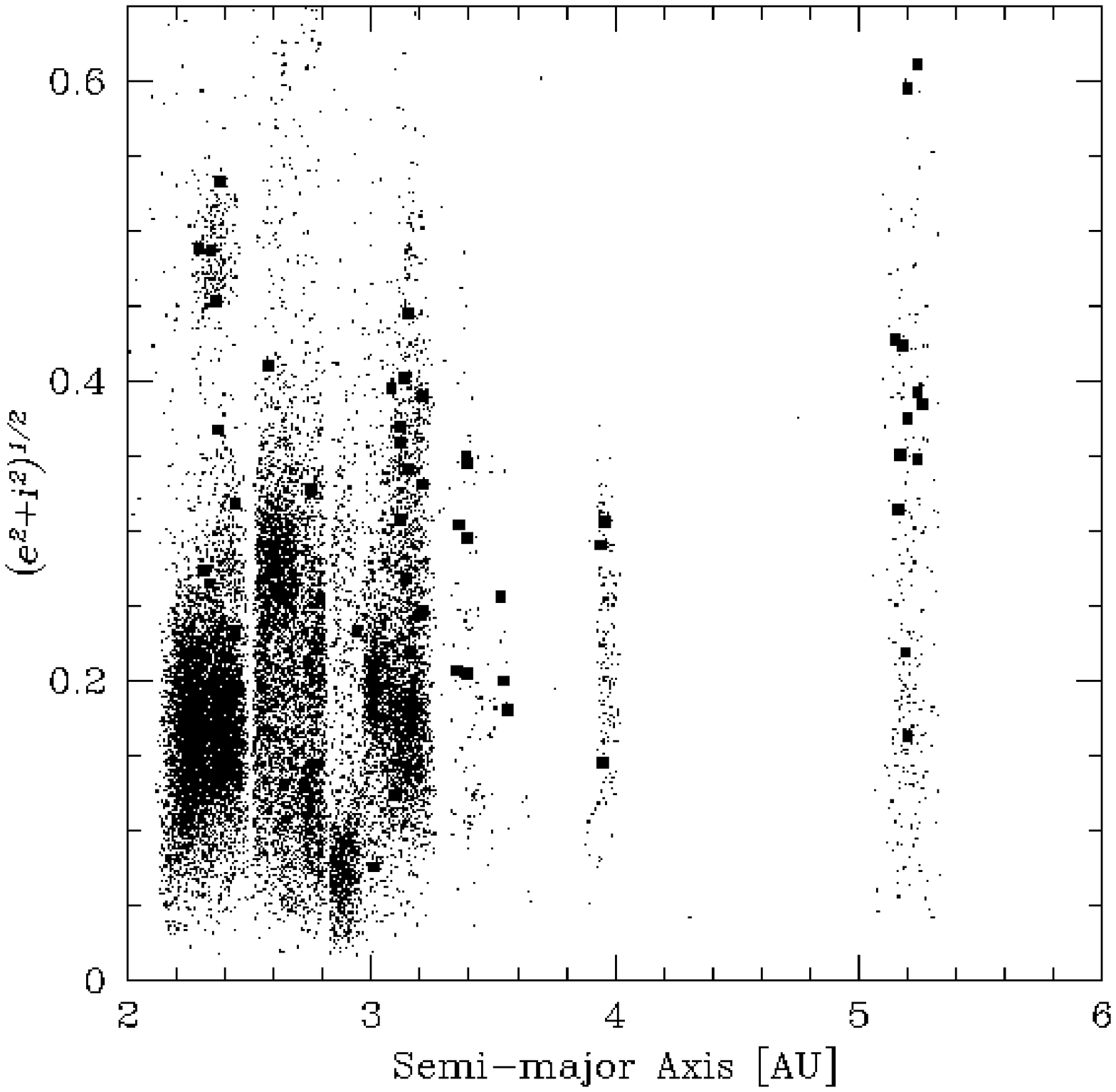}
 \caption{Distribution of asteroids. The vertical axis represents relative
velocity of the asteroids. The filled squares indicate D-type asteroids,
and the dots indicate asteroids other than D-type.
Data were compiled from \cite{Busb}, \cite{Lazzaro},
and this work. D-type asteroids in the inner main-belt region do not
have small relative velocity.}
 \label{dei}
\end{figure}

Figure \ref{dei} shows the inclination and eccentricity of asteroids as a 
function of the orbital semi-major axis.
The D-type asteroids show two distinct distribution.
One is the asteroids with the semi-major axis of 5.2 AU.
Those objects are called Trojan asteroids, resonant asteroids to Jupiter.
Another distribution is the D-type asteroids in the inner main-belt
but with the high relative velocity.
\cite{Carvano} classified asteroids based on the SDSS MOC 4 colors.
They indicated that the D-type asteroids are evenly distributed in the main-belt,
but those are not seen in the outer belt with the inclinations greater than 20 
\degr.
On the other hand, \cite{DeMeo2013} and \cite{DeMeo2014} found evidence
for the D-type asteroids in the inner- and mid-belts.
Our result is consistent with the results of \cite{DeMeo2013} and \cite{DeMeo2014}.

However, these results are not consistent with the influx of primitive material 
from Nice model like migration.
For example, \cite{Levison} investigated orbital migration of trans-Neptunian
objects.
They focused on bodies with the size larger than 40 km and
conclude that a significant fraction of objects in the main-belt are 
captured primordial trans-Neptunian objects.
They also find that the D-type and P-type material do not
come closer than 2.6 AU.
This result of the numerical simulation is not consistent with
the result of the observation in this work.

\cite{Walsh} proposed a dynamical model of the early Solar System,
so-called a "grand tack model".
In this model, Jupiter first migrated inward by gas drag of the 
protoplanetary disk.
When Jupiter reached at $\sim$ 1.5 AU from the Sun,
Saturn grew to 60 M$_{\earth}$, then migrated inward.
At the same time, Jupiter tacked and migrated outward.
The moving Jupiter initially scattered most of the planetesimals
in the main-belt, but then repopulates the main-belt,
where inner-belt asteroids originate between 1 and 3 AU
and outer-belt asteroids between and beyond the giant planets.
This distribution is consistent with the two separate populations
of the asteroids, if the planetesimals
from the inner disk are considered to be the S-type and those from 
the outer regions the C-type.
The model indicates that both types of the asteroids share similar 
distributions of eccentricity and inclination;
most of the outer disk objects on planet-crossing
orbits have high eccentricity, while many of the objects
from between the giant planets were scattered earlier and damped
to lower-eccentricity planet-crossing orbits.
We revealed the D-type asteroids in the inner main-belt 
but with high relative velocity.
Such a population is not predicted by \cite{Walsh}.

\cite{Ida} indicated that
the planetesimals were distributed in a 
V-shape with the apex at Jupiter in this diagram,
if they were formed near to Jupiter and were subsequently scattered.
For D-type asteroids with a semi-major axis smaller than 5 AU,
those asteroids with a smaller semi-major axis tend to have a large
$(e^{2}+i^{2})^{1/2}$ value.
This distribution of D-type asteroids might trace the left-side
of the V-shaped structure.
Further identification of D-type asteroids based on spectroscopic observations
is required to confirm the gravitational perturbation of Jupiter
on planetesimals.

\section{Conclusions}

We obtained low-resolution optical spectra of 51 asteroids.
Most of these were main-belt asteroids with a high orbital inclination.
We assigned spectral types to 38 asteroids. 
Eight asteroids were classified as D-type.
We confirmed the existence of the inner main-belt D-type asteroids.
We also revealed that the inner main-belt D-type asteroids have 
higher relative velocity.
\begin{enumerate}
\item For the asteroids with a semi-major axis of between 2.1 AU and 3.3 AU
and with a high orbital inclination, the fraction of D-type asteroids
was higher than that of the asteroids with the same range of semi-major axis
and with a low orbital inclination.
\item The fractions of C- and S-type asteroids are comparable between
those asteroids with a semi-major axis of between 2.1 AU and 2.5 AU
and a high orbital inclination and asteroids with
a low orbital inclination.
\item An abundant population of D-type main-belt asteroids with a high
orbital inclination indicates that a fraction of the high orbital
inclination asteroids were formed near Jupiter, and 
then migrated inward obtaining a highly-inclined orbit.
\end{enumerate}

\begin{acknowledgements}
We thank the members of staff and operators at the UH 2.2 m telescope
and the IGO 2 m telescope.
This research is partially supported by "the Japan-India Cooperative 
Science Program" carried out by Japan Society for the Promotion of 
Science (JSPS) and the Department of Science and Technology (DST), 
Government of India.
\end{acknowledgements}

\bibliographystyle{raa}
\bibliography{asteroid}

\end{document}